# The Jeffery-Hamel Flow and Heat Transfer of Nanofluids by Homotopy Perturbation Method and Comparison with Numerical Results

Majid Pourabdian[1], Mehran Qate[2], Mohammad Reza Morad[3], Alireza Javareshkian[4]

1,2,3-Department of Aerospace Eng., Sharif University of Technology, Tehran, Iran
4-Department of Aerospace Eng., Amirkabir University of Technology, Tehran, Iran

**Abstract**
This paper considers the influence of nanoparticles on the nonlinear Jeffery-Hamel flow problem. Investigation is performed for three types of nanoparticles namely copper Cu, alumina $Al_2O_3$ and titania $TiO_2$ by considering water as a base fluid. The resulting nonlinear governing equations and their associated boundary conditions are solved for both semi analytical and numerical solutions. The semi analytical solution is developed by using Homotopy Perturbation Method (HPM) whereas the numerical solution is presented by Runge-Kutta scheme. Dimensionless velocity, temperature, skin friction coefficient and Nusselt number are addressed for the involved pertinent parameters. It is observed that the influence of solid volume fraction of nanoparticles on the heat transfer and fluid flow parameters is more noticeable when compared with the type of nanoparticles. The achieved results reveal that HPM is very effective, convenient and accurate for this problem.

## 1. Introduction

Most of the scientific problems and phenomena are modeled by nonlinear ordinary or partial differential equations. In recent years, many powerful methods have been developed to construct explicit analytical solution of nonlinear differential equations. One of the most applicable semi analytical techniques is Homotopy Perturbation Method (HPM) which has been successfully applied to solve many types of nonlinear problems [1,2].

Jeffery [3] and Hamel [4] have worked on incompressible viscous fluid flow through convergent-divergent channels, mathematically. A survey of early information on this problem can be found in the refs. [5,6]. The Jeffery-Hamel flow problem is solved by other techniques such as perturbation method [7], the domain decomposition method [8,9], the homotopy analysis method [10] and the spectral-homotopy analysis method [11].

Recently, there is an increasing interest of the researchers in the analysis of nanofluids. The word nanofluid was firstly introduced by Choi [12]. Bachok and Ishak, [13], studied numerically flow and heat transfer characteristics of a nanofluid over a moving plate and reported that incorporating nanoparticles in the base fluid water results in enhancing the friction coefficient and Nusselt number and respectively thermal conduction. Kuznetsov and Nield, [14], studied the influence of nanoparticles on natural convection boundary layer flow past a vertical plate by taking Brownian motion and thermophoresis into account. Nadeem et al, [15], investigated HAM solutions for boundary layer flow in the region of the stagnation point towards a stretching sheet. Rana and Bhargava, [16], studied the effect of Brownian motion and thermophoresis on natural convection of a nanofluid over a nonlinearly stretching sheet by means of numerical methods. Moradi, [17], investigated the heat transfer and viscous dissipation effects on the Jeffery-Hamel nanofluids using differential transformation method (DTM).

In present research, the HPM is applied to find the analytical solutions of nonlinear differential problems governing Jeffery-Hamel flow with respect to the heat transfer and viscous dissipation in nanofluids. The outline of the paper is as follows: In Section 2, we describe the problem and its mathematical analysis. Section 3 expresses the basic concepts of Homotopy Perturbation Method and the obtained solution and results are presented in Section 4. Finally, some conclusions are given in Section 5.

## 2. Problem Statement and Mathematical Analysis

Consider the steady two-dimensional flow of an incompressible conducting viscous fluid between two rigid plane walls that meet at an angle 2α in a water-base nanofluid containing different types of nanoparticles namely Cu, $Al_2O_3$ and $TiO_2$ as shown in Fig. 1. We assume that the velocity is purely radial and depends on *r* and *θ* such that the velocity *V= (u(r,θ),0)*. The equations of continuity, motion and energy considering viscous dissipation for the problem under consideration give:

$$\frac{\rho_{nf}}{r}\frac{\partial}{\partial r}\bigl(ru(r,\theta)\bigr) = 0 \qquad (1)$$

$$u(r,\theta)\frac{\partial u(r,\theta)}{\partial r} = -\frac{1}{\rho_{nf}}\frac{\partial p}{\partial r} + \frac{\mu_{nf}}{\rho_{nf}}\Bigl[\frac{\partial^2 u(r,\theta)}{\partial r^2} + \frac{1}{r}\frac{\partial u(r,\theta)}{\partial r} + \frac{1}{r^2}\frac{\partial^2 u(r,\theta)}{\partial \theta^2} - \frac{u(r,\theta)}{r^2}\Bigr] \qquad (2)$$

$$-\frac{1}{\rho_{nf} r}\frac{\partial p}{\partial \theta} + \frac{2\mu_{nf}}{\rho_{nf} r^2}\frac{\partial u(r,\theta)}{\partial \theta} = 0 \qquad (3)$$

1. M.Sc. Student, 09195524354, m.pourabdian@gmail.com (corresponding author)
2. M.Sc. Student
3. Assistant Professor
4. B.Sc.

$$u(r,\theta)\frac{\partial T(r,\theta)}{\partial r} = \alpha_{nf}\left(\frac{\partial^2 T(r,\theta)}{\partial r^2} + \frac{1}{r}\frac{\partial T(r,\theta)}{\partial r} + \frac{1}{r^2}\frac{\partial^2 T(r,\theta)}{\partial \theta^2}\right) + \frac{\mu_{nf}}{(\rho C_p)_{nf}}\left(4\left(\frac{\partial u(r,\theta)}{\partial r}\right)^2 + \frac{1}{r^2}(\frac{\partial u(r,\theta)}{\partial \theta})^2\right) \quad (4)$$

with the subjected boundary conditions:
At the channel centerline: $\frac{\partial u(r,\theta)}{\partial \theta} = 0, \frac{\partial T}{\partial \theta} = 0,$ $u(r,\theta) = U$
At the plates, making the body of the channel:
$u(r,\theta) = 0, T = T_w$
Here, $\alpha_{nf}$ is the thermal diffusivity of the nanofluid and $\rho_{nf}$ is the density of the nanofluid, and $\mu_{nf}$ is the viscosity of the nanofluid which are given by Oztop and Abu-Nada, [18]:

$$\rho_{nf} = (1-\varphi)\rho_f + \varphi\rho_s, \quad \mu_{nf} = \frac{\mu_f}{(1-\varphi)^{2.5}} \quad (5)$$

The value of $\alpha_{nf}$ is [11]:

$$\alpha_{nf} = \frac{k_{nf}}{(\rho C_p)_{nf}}, (\rho C_p)_{nf} = (1-\varphi)(\rho C_p)_f + \varphi(\rho C_p)_s,$$
$$\frac{k_{nf}}{k_f} = \frac{(k_s + 2k_f) - 2\varphi(k_f - k_s)}{(k_s + 2k_f) + 2\varphi(k_f - k_s)} \quad (6)$$

where, $\varphi$ is the nanoparticle volume fraction, $(\rho C_p)_{nf}$ is the heat capacity of the nanofluid, $k_{nf}$ is the thermal conductivity of the nanofluid, $k_f$ and $k_s$ are the thermal conductivities of the base fluid and of the solid fractions, respectively, and $\rho_f$ and $\rho_s$ are the densities of the base fluid and of the solid fractions, respectively as mentioned in Bachok, [13]. The use of the above expression for $k_{nf}/k_f$ is restricted to spherical nanoparticles, [18,19]. Also, the viscosity of the nanofluid $\mu_{nf}$ has been approximated by Brinkman, [20]. Equation (1) yields

$$f(\theta) = ru(r,\theta) \quad (7)$$

we introduce

$$f(\eta) = \frac{f(\theta)}{f_{max}}, \quad f_{max} = rU, \quad \eta = \frac{\theta}{\alpha}, \quad \xi(\eta) = \frac{T}{T_w} \quad (8)$$

now we eliminate $p$ between Eqs. (2) and (3), we arrive to the following equations:

$$f'''(\eta) + 2\alpha Re\left[(1-\varphi)^{2.5}\left(1-\varphi+\varphi\frac{\rho_s}{\rho_f}\right)\right]f(\eta)f'(\eta) + 4\alpha^2 f'(\eta) = 0 \quad (9)$$

$$\frac{1}{[1-\varphi+\varphi(\rho c_p)_s/(\rho c_p)_f]}\left[\frac{k_{nf}}{k_f}\xi'' + \frac{Pr\,Ec}{(1-\varphi)^{2.5}}(4\alpha^2 f^2 + f'^2) = 0 \quad (10)$$

with the following boundary conditions:

$$f(0)=1, f'(0)=0, f(1)=0 \quad (11)$$
$$\xi(1)=1, \xi'(0)=0 \quad (12)$$

where the Reynolds number $Re$, the Eckert number $Ec$ and Prandtle number $Pr$ are expressed as:

$$Re = \frac{f_{max}\rho_f\alpha}{\mu_f} = \frac{U_{max}r\rho_f\alpha}{\mu_f}, Pr = \frac{\mu_f(C_p)_f}{k_f}, Ec = \frac{U^2}{T_w(C_p)_f} \quad (13)$$

skin friction coefficient ($C_f$) and shear stress ($\tau_w$) expressions are defined:

$$C_f = \frac{\tau_w}{\rho_f U_{max}^2}, \tau_w = \mu_{nf}\left(\frac{1}{r}\frac{\partial u(r,\theta)}{\partial \theta}\right) \quad (14)$$

If we substitute Eq. (8) into Eq. (14), the skin friction coefficient is:

$$C_f = \frac{1}{Re(1-\varphi)^{2.5}}f'(1) \quad (15)$$

The local Nusselt number $Nu$ and heat transfer rate are:

$$Nu = \frac{rq_w|_{\theta=\alpha}}{k_f T_w}, q_w = -k_{nf}\nabla T \quad (16)$$

The above equation in view of Eq. (8) yields:

$$Nu = -\frac{1}{\alpha}\frac{k_{nf}}{k_f}\xi'(1) \quad (17)$$

The equations derived in this section were a mathematical model to reach to nonlinear differential equations along with the boundary conditions.

## 3. Solution by Homotopy Perturbation Method

In this study, we apply the homotopy perturbation method (HPM) which is a strong tool to solve partial differential equations to the discussed problem. To explain this method, let us consider the following general nonlinear equation

$$A(u) - f(r) = 0 \qquad r \in \Omega \quad (18)$$

with the boundary condition of:

$$B(u, \frac{\partial u}{\partial n}) = 0 \qquad r \in \Gamma \quad (19)$$

where $A$ is a general differential operator, $B$ is a boundary operator, $f(r)$ is a known function and $\Gamma$ is the boundary of the domain $\Omega$. The operator $A$ can divided in two parts $L$ and $N$, where $L$ is linear, and $N$ is nonlinear, therefore Eq. (18) can be written as

$$L(u) + N(u) - f(r) = 0 \quad (20)$$

By using homotopy technique, one can construct a homotopy $v(r,p) : \Omega \times [0, 1] \rightarrow R$ which satisfies homotopy equation

$$H(v,p) = L(v) - L(u_0) + p\,L(u_0) + p[N(v) - f(r)] = 0 \quad (21)$$

or

$$H(v,p) = (1-p)\,[L(v) - L(u_0)] + p\,[A(v) - f(r)] = 0 \quad (22)$$

where $p \in [0, 1]$ is an embedding parameter, and $u_0$ is the first approximation that satisfies the boundary conditions. Clearly, we have

$$H(v,0) = L(v) - L(u_0) = 0, H(v,1) = A(v) - f(r) = 0 \quad (23)$$

The changing process of p from zero to unity is just that of $v(r,p)$ changing from $u_0(r)$ to $u(r)$. If, the embedding parameter $p$ $(0 \leq p \leq 1)$ is considered as a "small parameter", applying the classical perturbation technique, we can naturally assume that the solution of Eqs. (21) and (22) can be given as a power series in $p$, i.e.,

$$v = v_0 + pv_1 + p^2 v_2 + ... \quad (24)$$

and setting p=1 results in the approximate solution of Eq. (25) as

$$u = \lim_{p \to 1} v = v_0 + v_1 + v_2 + \cdots \quad (25)$$

The convergence of series (25) has been proved He in his paper [1] and the number of terms in the series have to be as much as required convergence.

### 4. Results and Discussion

Semi analytical solution of the nonlinear differential governing equations (9) and (10) with the specified boundary conditions (11) and (12) obtained using HPM method. We have also employed fourth-order Runge-Kutta method to solve the problem numerically. The investigation has been carried out for three different types of nanoparticles where the thermophysical properties of these nanoparticles are shown in table 1.

**Table 1 The physical properties of nanofluids and base fluid [18]**

| Physical Properties | Water (base fluid) | Cu | $Al_2O_3$ | $TiO_2$ |
|---|---|---|---|---|
| $\rho(kg/m^3)$ | 997.1 | 8933 | 3970 | 4250 |
| $C_p(J/kgK)$ | 4179 | 385 | 765 | 686.2 |
| k(W/mK) | 0.613 | 400 | 40 | 8.9538 |

Firstly, it is compared the accuracy of HPM method with numerical and other reported results [11, 20] for $Re=50, \alpha=5°$ in table 2. It is transparent that the HPM results are in an excellent agreement with OHAM, SHAM and numerical results.

**Table 2 The results of HPM, OHAM, SHAM and numerical solution for f(η) when Re=50 and α=5°**

| η | HPM | OHAM[21] | SHAM[11] | Numerical |
|---|---|---|---|---|
| 0 | 1 | 1 | 1 | 1 |
| 0.1 | 0.9824314771 | 0.98251808 | 0.982431 | 0.982431 |
| 0.2 | 0.9312268428 | 0.93156588 | 0.931226 | 0.931226 |
| 0.3 | 0.8506123257 | 0.8513815 | 0.850611 | 0.850611 |
| 0.4 | 0.7467931374 | 0.74826039 | 0.746791 | 0.746792 |
| 0.5 | 0.6269505503 | 0.62953865 | 0.626848 | 0.6268488 |
| 0.6 | 0.4982362037 | 0.50242894 | 0.498234 | 0.498234 |
| 0.7 | 0.3669671316 | 0.37293383 | 0.366966 | 0.366966 |
| 0.8 | 0.2381237540 | 0.24508197 | 0.238124 | 0.238124 |
| 0.9 | 0.1151516618 | 0.1207156 | 0.115152 | 0.115152 |
| 1 | 0 | 0.000000001 | 0 | 0 |

Figure 2-4 show the influence of solid volume fraction on the normalized velocity profile when Re=50, α=5° for Cu, $Al_2O_3$, $TiO_2$, respectively. It can be seen from Fig. 2 that velocity decreases when the solid volume fraction increases for Cu nanoparticle. However, the velocity increases when the solid volume fraction increases for $Al_2O_3$ and $TiO_2$ nanoparticles (see Figs. 3, 4). The obvious agreement between HPM and numerical results are noticeable in Figs. 2-4 as well. Moreover, it is worth to mention that the impact of solid volume fraction in Cu-water nanofluid is more evident than the other nanoparticles.

The changes in the considered nanoparticles on the normalized velocity when $Re=70, \alpha=5°$ and $\varphi=0.2$ is displayed in Fig. 5. It is found that the values of velocities for $Al_2O_3$ and $TiO_2$ are almost the same and larger than Cu nanoparticle velocity. Figure 6 illustrates the normalized temperature profile with different types of solid volume fractions for water-Cu nanofluid when $Re=50, Ec=0.5$ and $\alpha=5°$. It is observed that the temperature increase with the increasing of solid volume fraction. Prandtle number is considered to be 6.2 in the whole solution. The temperature of different nanoparticles is shown in Fig. 7. It is clear that Cu nanoparticle has higher temperature than the other nanoparticles.

The variation of skin friction coefficient and Nusselt number with Reynolds number in three types of nanoparticles for different solid volume fraction are shown in Figs. 8 and 9, respectively. It is observed that both skin friction coefficient and Nusselt number decrease with increasing of Reynolds number. Further, when solid volume fraction increases, the skin friction coefficient and Nusselt number increase as well. Besides, the values of skin friction coefficient and Nusselt number for $Al_2O_3$ are larger than the other nanoparticles.

### 5. Conclusion

It the present paper, we investigated the influences of nanofluid and heat transfer effects on the quantities of the Jeffery-Hamel Flow. The governing nonlinear partial differential equations are solved by HPM method and numerically. It is observed that the effect of solid volume fraction of nanoparticles on the heat transfer and fluid flow parameters is more pronounced when compared with the type of nanoparticles. It is also found that skin friction coefficient and Nusselt number for $Al_2O_3$ nanofluid is the highest in comparison to the other two nanoparticles. The obtained results show that Homotopy Perturbation Method is a very useful, convenient and practical technique to get highly accurate solution to such kind of nonlinear problems.

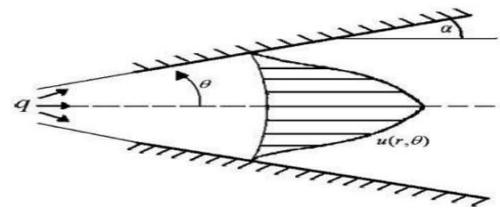

**Figure 1 Geometry of the problem [17]**

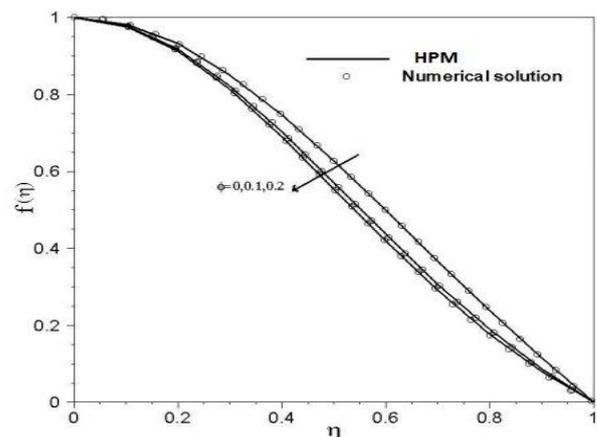

**Figure 2** Normalized velocity profile with different types of solid volume fraction for water-Cu nanofluid when Re=50 and $\alpha=5°$

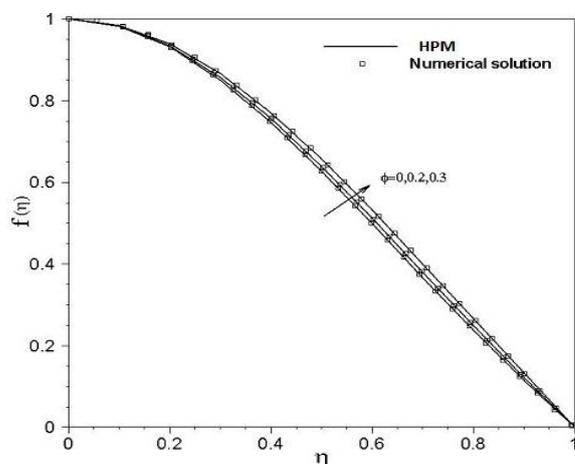

**Figure 3** Normalized velocity profile with different types of solid volume fraction for water-$Al_2O_3$ nanofluid when Re=50 and $\alpha=5°$

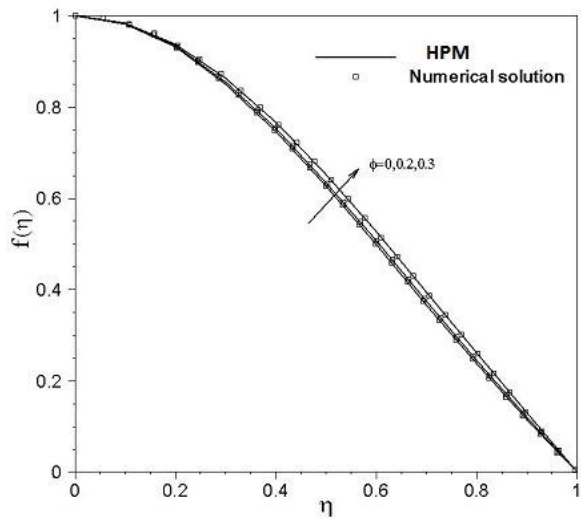

**Figure 4** Normalized velocity profile with different types of solid volume fraction for water-TiO$_2$ nanofluid when Re=50 and α=5°

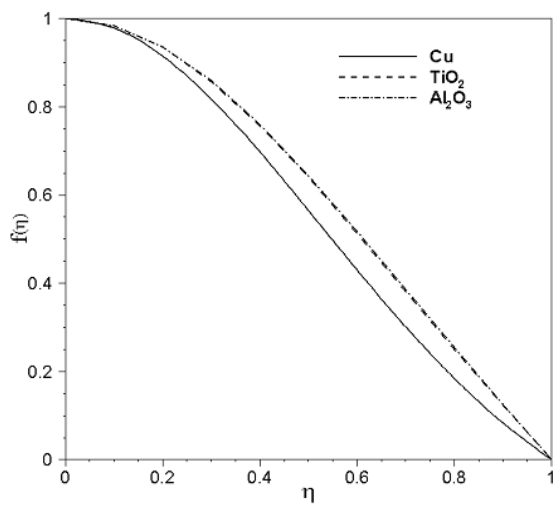

**Figure 5** Normalized velocity profile for three types of nanoparticles when Re=70, α=5° and φ=0.2

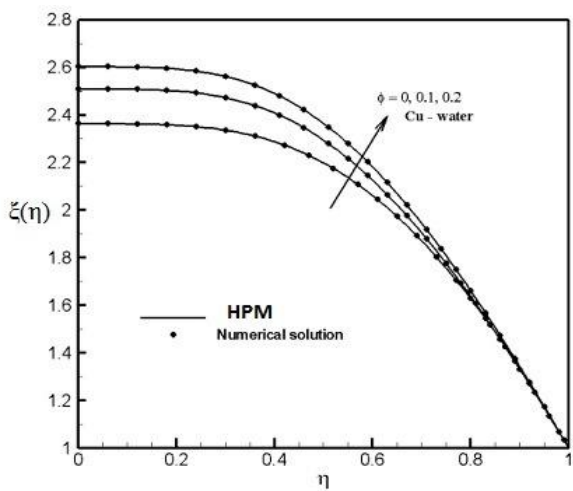

**Figure 6** Normalized temperature profile with different types of solid volume fraction for water-Cu nanofluid when Re=50, Ec=0.5 and α=5°

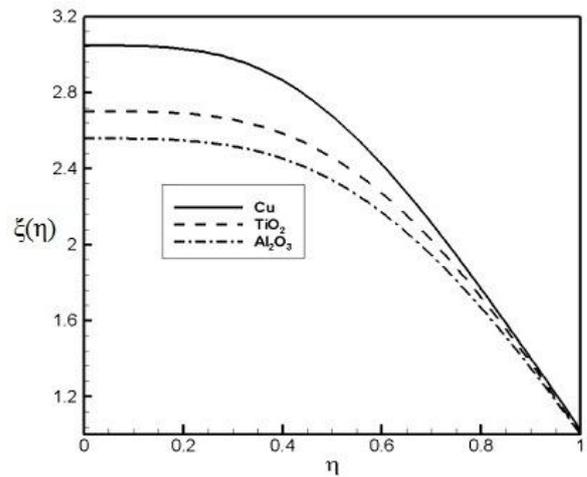

**Figure 7** Normalized temperature profile for three types of nanoparticles when Re=50, Ec=0.5 and α=5°

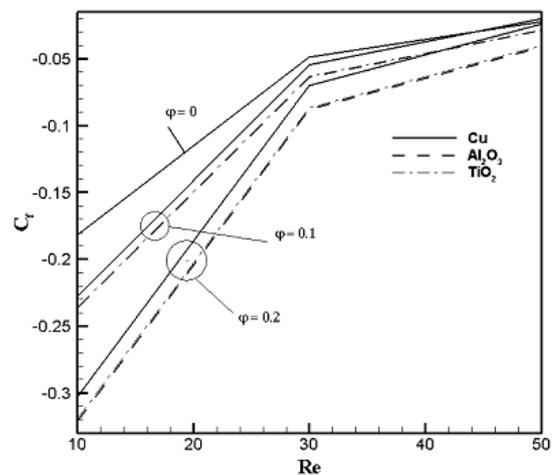

**Figure 8** The diagram of skin friction coefficient with Reynolds number in three types of nanoparticles for different solid volume fractions

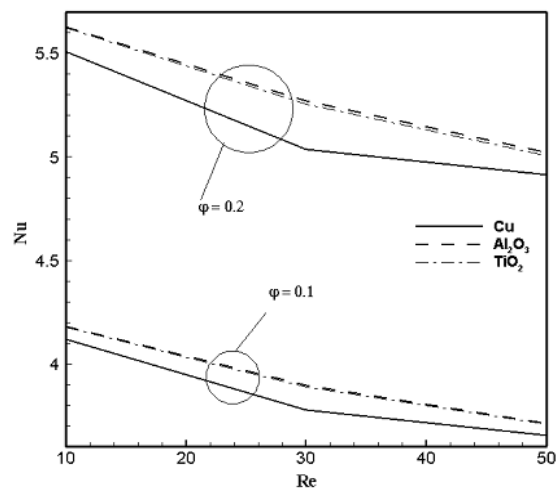

**Figure 9** The diagram of skin Nusselt number with Reynolds number in three types of nanoparticles for different solid volume fractions